# Beyond surveys: A High-Precision Wealth Inequality Mapping of China's Rural Households Derived from Satellite and Street View Imageries

Weipan Xu[1], Yaofu Huang[2], Qiumeng Li[3], Yu Gu[4], Xun Li[5*]

Affiliations:

1. Sannong Data (Guangzhou) Inc., Guangzhou, China (xuweipan@mail2.sysu.edu.cn )
2. School of Geography and Planning, Sun Yat-sen University, Guangzhou, China
3. The Hong Kong University of Science and Technology (Guangzhou), Guangzhou, China
4. School of Geography and Planning, Sun Yat-sen University, Guangzhou, China
5. School of Geography and Planning, Sun Yat-sen University, Guangzhou, China (lixun@mail.sysu.edu.cn )

**Abstract**

Wide coverage and high-precision rural household wealth data is an important support for the effective connection between the national macro rural revitalization policy and micro rural entities, which helps to achieve precise allocation of national resources. However, due to the large number and wide distribution of rural areas, wealth data is difficult to collect and scarce in quantity. Therefore, this article attempts to integrate "sky" remote sensing images with "ground" village street view imageries to construct a fine-grained "computable" technical route for rural household wealth. With the intelligent interpretation of rural houses as the core, the relevant wealth elements of image data were extracted and identified, and regressed with the household wealth indicators of the benchmark questionnaire to form a high-precision township scale wealth prediction model (r=0.85); Furthermore, a national and township scale map of rural household wealth in China was promoted and drawn. Based on this, this article finds that there is a "bimodal" pattern in the distribution of wealth among rural households in China, which is reflected in a polarization feature of "high in the south and low in the north, and high in the east and low in the west" in space. This technological route may provide alternative solutions with wider spatial coverage and higher accuracy for high-cost manual surveys, promote the identification of shortcomings in rural construction, and promote the precise implementation of rural policies.

**Keywords:** wealth inequality; Rural houses; Village street view; Remote sensing images; Computability

## 1 Introduction

In pursuit of rural revitalization, China targets the unbalanced and insufficient development of rural areas, necessitating studies on rural wealth differentiation and precise policy application as a core geographical concern[1]. With the increasing complexity of rural dynamics, fine-grained data becomes indispensable in formulating responsive policies and in assessing their impact on enhancing the well-being of rural populations, both currently and for the future[2,3]. Nonetheless, comprehensive micro-level spatial data delineating rural wealth is a rarity. For instance, the "Urban and Rural Construction Statistical Yearbook" uncovers a population decline alongside an increase in housing area, indicating substantial wealth redistribution within the rural expanse [4]. But at a village or household level, such trends remain opaque.

The reliance on survey methodologies to chart rural wealth has been met with limitations in scope, high costs, and temporal challenges, particularly in remote regions [5]. Even decadal census efforts fall short in dynamism and granularity, rendering them insufficient for the nuanced study of rural wealth, where farmers' non-wage incomes dominate yet are often occluded by a lack of detailed data [6]. This signals the need for a scalable, cost-efficient, and detailed data framework to complement and enhance survey approaches.

Agricultural housing stands out as the primary spatial asset for rural wealth, reflecting the investment patterns of farmers and their aspirations for improved living [7]. Over decades, changes in housing—from simple cottages to multi-story buildings—mirror economic progress and the increasing weight of real estate within rural assets, climbing from 33% to 60% of household wealth from 1990 to 2013 [8]. The wealth indices derived from agricultural housing features have gained traction for their interpretative strength, with a proven global and continental applicability [9–12].

The potential of remote sensing imagery to economically scale wealth prediction has been recognized, with nighttime light data initially serving as a socioeconomic proxy [13–15]. Yet, its efficacy is inconsistent across scales. Daytime images, enriched with discernible wealth-related features, offer an enhanced predictive edge as supported by notable publications and subsequent wealth distribution studies [9–12]. The progression into sub-meter resolution has further refined spatial data for housing wealth assessment, as exemplified by the China agricultural housing vector database [16,17].

In a three-dimensional extension, crowdsourced street view data offers a richer tapestry of human activity and wealth indicators—providing details often missed by top-down remote sensing. This method, bolstered by the widespread use of smartphones and social media in rural regions, furnishes an abundance of grassroots data, from housing appearances to material quality [18–20].

This paper synthesizes remote sensing with street view data, centering on agricultural housing to chart a national, township-level wealth index, revealing the contours of wealth disparity. We assess over 1.85 million street views, constructing a model that marries subjective quality evaluation with objective housing data—dimensions, materials, and amenities—to profile agricultural housing wealth. Incorporating these features with survey-derived wealth labels, we pioneer a machine learning model that fine-tunes the prediction of wealth indicators and indices, ultimately mapping China's rural wealth landscape at an unprecedented township granularity.

The study's culmination provides a wealth distribution map that speaks directly to policy-makers, equipping them with the precise data necessary for targeted rural policy formulation and effective revitalization initiatives. It's an endeavor that not only maps wealth but also the pathways to a sustainable rural future, in alignment with the global Sustainable Development Goals.

## 2 Methods

This paper first establishes an actual wealth label database based on rural household questionnaire survey data and establishes a regression relationship with the feature characteristics of intelligent image interpretation to form a wealth prediction model. Furthermore, this paper extends the wealth prediction model to areas where questionnaires have not been conducted, based on large-scale remote sensing imagery and village street view data, forming a township-scale wealth distribution map of China. In terms of actual wealth label generation, this paper uses the rural household questionnaire database from the 2022 national rural construction evaluation, covering 28 provinces, 124 counties, 1,678 townships, and 180,000 questionnaires. These questionnaire data have a wide spatial coverage and a large volume, recording information on farmers' income, property, consumption, etc., and involving counties with different natural geographic conditions and development levels, which have high statistical representativeness. For feature input, the paper collected about 1.85 million agricultural housing street view images from the "Village Photo" crowdsourced mini-program and rural human settlement environment survey database for intelligent interpretation, combining the features of the existing agricultural housing vector database to train a wealth prediction model for multiple wealth indicators. Based on this, the paper first reveals the regional differences in the wealth of rural households in China at a national scale, using townships as units. Finally, the paper conducts an in-depth study on the regional differences in rural household wealth in China, studying the impact of natural and economic factors on wealth accumulation through regression analysis.

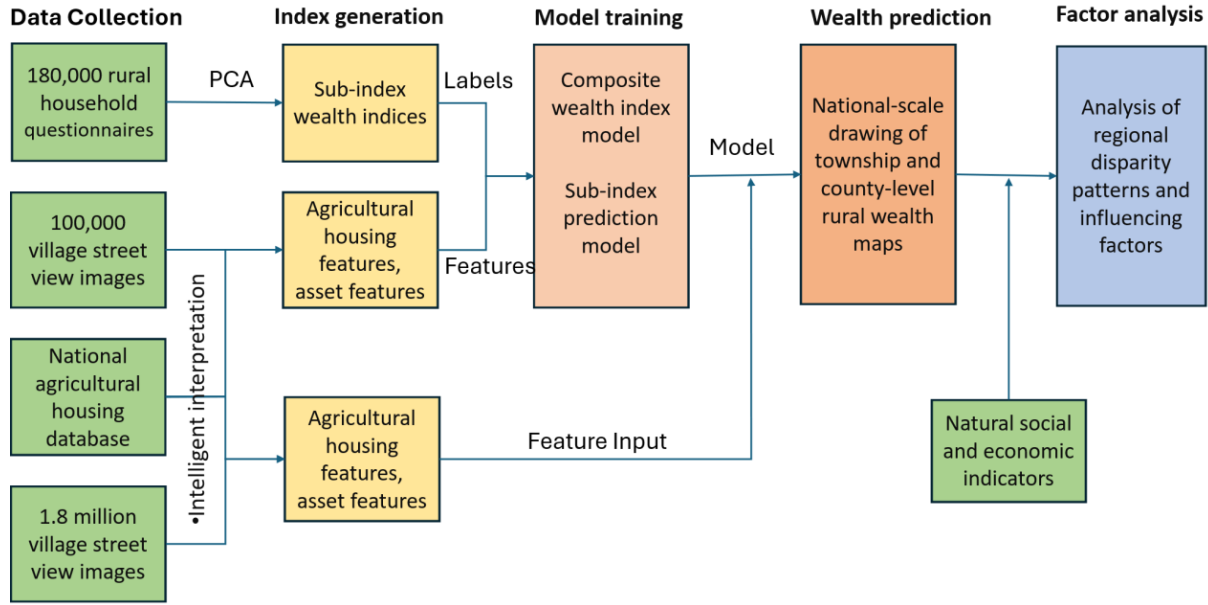

Figure 1. Comprehensive Measurement Process of Rural Household Wealth in China.

## 2.1 Research Data

The research data of this paper mainly covers three aspects: First, rural household questionnaire survey data, as a measure of actual rural wealth, used for label input in the prediction model. Second, village street views and remote sensing interpretation of the agricultural housing database, used for feature input in the prediction model. Among them, part of the data corresponding to the questionnaires is used to train the machine learning model, and the imagery data from areas not covered by the questionnaires is used to predict wealth distribution nationwide. Third, statistical data on natural, social, and economic aspects, used to analyze the factors affecting the regional distribution of wealth in China.

### 2.1.1 Village Questionnaire Survey for Generating Wealth Labels

The village questionnaire survey of rural construction evaluation is an important basis for the wealth labels in this paper. In 2022, the evaluation covered 102 national sample counties and 22 provincial sample counties, totaling 124 sample counties. These sample counties have a wide coverage and strong regional representativeness. When selecting national sample counties, 3 to 5 counties from each province are selected to report, considering different regional environmental backgrounds and economic development levels. Therefore, using the sample counties from the rural construction evaluation as data labels ensures the typicality and representativeness of label input and aids in extending the prediction model nationwide. The paper collected about 180,000 valid questionnaires, covering 1,678 townships. The content of the questionnaires mainly involves family basic information, agricultural housing and supporting facilities, and includes rich information on wealth elements such as income, assets, and consumption. The volume of questionnaire data is sufficient, and the content dimensions are rich, which helps to train a high-performance model.

Wealth is commonly viewed as an asset. From a dynamic perspective, income can be converted into savings and solidify into stock assets, while consumption is the expenditure of stock assets [21]. Therefore, this paper proposes to assess the wealth level of rural households from three dimensions: income, assets, and consumption. There are a total of 7 related questions in the villagers' questionnaire, as shown in Table S3 in the Supplementary Materials (SM). In terms of income, since income information is relatively private, the questionnaire is designed to use ranges as answer options. Although setting income intervals reduces precision, it can better avoid the issue of respondents concealing information. In terms of assets, core assets such as agricultural housing, home appliances, and cars are mainly considered. Among them, agricultural housing is the most important asset for farmers, accounting for nearly 60% of household wealth. Cars are also one of the more important assets. In terms of consumption, electricity consumption is mainly considered. The higher the wealth level, the higher the electricity consumption tends to be.

### 2.1.2 Village Street View and Agricultural Housing Database for Generating Wealth Features

Village street views primarily come from the "Village Photo" rural graphic crowdsourcing platform. It supports the collection of human settlement environmental element pictures and questionnaire survey work for the corresponding sample counties in the form of public welfare, accumulating a large number of village street view pictures spatially matched with the villager questionnaire data. There are approximately 100,000 street view pictures of villages with agricultural housing as the main subject, which can completely cover the 1,678 townships mentioned in the questionnaire data. This enables the questionnaire data, which serve as wealth labels, to be associated with the village street views generating wealth features. For model extension and prediction, this paper used the village human settlement environment survey database. There are about 1.85 million village street views, covering 30,667 townships (streets) in China, accounting for about 75% of the national total. Based on this, the paper can draw a national, township-scale map of rural household wealth.

Additionally, this paper used the vector boundary map of Chinese townships to conduct spatial statistical analysis of the Chinese agricultural housing vector map, which includes area and age information. Through this analysis, indicators such as the number of agricultural houses in each township, average building area, and the proportion of old-style agricultural houses were calculated.

## 2.2 Wealth Measurement Methods

### 2.2.1 Real Wealth Labels Based on Villager Questionnaires

This paper calculates and obtains related wealth indicators from the rural household questionnaire data. Starting from three dimensions - income, assets, and consumption - a total of 13 sub-indexes were obtained. In addition, principal component analysis was used to reduce dimensions and aggregate these 13 sub-indexes, with the first principal component being used as a composite wealth index. Both these sub-indexes and the composite index were used for model training.

**Sub-indexes.** A series of characteristics such as rural housing, living facilities, and energy consumption are closely related to the wealth of farmer households. Especially for rural housing, housing construction is an important indicator of rural economic development, with more than 60% of farmers' income being invested in their homes. As income levels increase, cars also become a major tool for wealthier farmers to enjoy a better life, which in turn indirectly reflects the wealth level of farmer households. Therefore, in the 13 sub-indexes of this paper, housing-related indicators make up the majority, totaling 10 items. The rest include family income, car ownership, and electricity consumption indicators. The specific indicators are defined in Table S4. Housing-related indicators mainly consider factors such as floor height, base area, house structure, exterior wall decoration, toilet, kitchen, bathroom, and internal facilities such as tap water, air conditioning, broadband, etc. Farmers invest their income in upgrading their homes to achieve their ideal comfortable living environment. From the basic size and safety of the house to the aesthetic appearance and the comfort and convenience of facilities, these are all reflections of the gradual increase in farmers' wealth accumulation levels. Clearly, larger houses and more robust structures require more capital investment, and the same goes for purchasing more air conditioning, broadband, and other facilities. In addition, the functional zoning within the house and comprehensive construction are one of the characteristics of modernized agricultural housing, where independent kitchens, bathrooms, and flush toilets are also important guarantees of housing quality. However, the provision of these facilities is still not widespread in many backward areas. For example, only 54% of self-built homes nationally are equipped with flush toilets (Table S4), and some areas still use pit latrines. Cars are another important asset for farmer households. According to the questionnaire survey, about 62% of farmer households in China own a car. Since farmers usually count the cars of family members working away from home and not just those residing in the village, this percentage may be an overestimate of local rural assets. The average monthly electricity bill in the summer reflects the farmers' capacity to use household appliances and is another important representation of wealth level.

**Composite Index.** This paper further aggregates the above 13 indicators to form a composite wealth index. The method of principal component analysis is used to reduce the dimensions of the 13 indicators, and the first principal component is used as the composite wealth index. This type of operation is widely used in the process of composite wealth analysis to generate reliable and stable labels for machine learning.

For the combination of economic variables related to family wealth, the first principal component obtained through principal component analysis can often be interpreted as a family wealth index, or a proxy variable for the overall and long-term socioeconomic level of a family. When dimensionality reduction is performed on a set of related household economic variables using principal component analysis, the first principal component often captures the largest variance in the data. Since these economic variables all reflect family wealth, such as housing and cars, the first principal component is likely to represent a common wealth factor that explains the most significant variance in the data. In fact, from an economic sense, this implies an assumption: wealth is the most important factor affecting the differentiation of these economic variables. Therefore, it is reasonable to interpret the first principal component as a wealth index or a proxy variable for socioeconomic status.

$$pc_1 = w_1 x_1 + w_2 x_2 + \cdots + w_n x_n \tag{1}$$

From equation (1), in principal component analysis, each principal component is a linear combination of the original variables, where each variable is multiplied by its corresponding weight. These weights, also known as loadings or coefficients, are different for each variable in each principal component. Suppose there are n related variables X1, X2, ..., Xn. The first principal component PC1 can be expressed as: (1) Here, a1, a2,..., an are the weights assigned to each variable X1, X2,..., Xn in PC1. These weights determine the contribution of each variable to PC1. The selection of weights aims to maximize the variance explained by PC1 while ensuring that the first principal component is uncorrelated with the other principal components.

### 2.2.2 Wealth Feature Indicators Based on Intelligent Interpretation

This paper further extracts related wealth proxy variables from national remote sensing images and street views, realizing low-cost, large-scale, and high-precision feature extraction of rural household wealth elements (see S1 of Supplementary Materials). A total of 10 major wealth feature characteristics were identified, mainly including housing features and facility features. The specific feature indicator definitions and data sources are as follows in Table S5.

In the housing dimension, this paper primarily identified features such as the base area of agricultural housing, number of floors, exterior wall decoration types, and subjectively scored the quality of agricultural housing, while also counting the number of houses in townships. These variables collectively reflect the scale and quality of agricultural housing. The base area of the building and the number of floors complement each other, together representing the scale of the agricultural housing. The larger the base area and the higher the number of floors, the larger the scale of agricultural housing construction. The scale of agricultural housing is directly linked to the cost of raw construction. The costs of construction materials and labor such as steel, cement, and bricks are basically proportional to the size of the agricultural housing. Therefore, the larger the agricultural housing, the more wealth is embedded in it.

The type of exterior wall decoration further reflects the financial strength of the farmer. Houses with tiled exteriors are more aesthetically pleasing and sought after by many modern agricultural housing constructions, and the material cost is also higher, hence farmers who opt for tiles usually have more substantial financial resources. In contrast, houses with exposed exterior walls indicate that the farmer does not have sufficient funds to build a better house, reflecting a lower level of wealth accumulation. This paper also uses the quality of agricultural housing as one of the features reflecting the wealth of farmers. The quality of agricultural housing has a strong correlation with the disposable income of farmers. Families with higher income levels are likely to build higher-quality houses. Therefore, the quality of agricultural housing can be used to predict wealth levels. The total number of houses in townships reflects the aggregation degree of housing in the township and may produce a scale effect on the accumulation of farmers' wealth.

In the facilities dimension, this paper primarily identified facilities such as cars, motorcycles, and air conditioners. Cars are the second most important asset after housing. A car, along with a house, is an important representation of the economic status of a rural family. Especially in the rural marriage market, farmer families with cars often have a competitive edge. In rural areas, the presence and number of cars can also reflect the wealth level of farmers and villages. Motorcycles are also an important asset in the daily life of farmers but are slightly less indicative of wealth than cars.

Air conditioning is another important support facility for modern rural life. With global warming and an increase in extreme weather events, air conditioners have become an important guarantee for farmer families to maintain a comfortable living environment and combat heat or severe cold. Operating air conditioners also consumes a large amount of electricity. Air conditioning is not a rigid need, and families that can afford air conditioners typically have a not insignificant income. Additionally, since the air conditioning unit is usually mounted on the exterior wall, it is easily captured by village street views. Therefore, this paper conjectures that air conditioning can be used to predict the wealth level of farmers.

After identifying or extracting the above feature elements, this paper calculates the average value or proportion of the related indicators at the township level. Ideally, wealth-related variables should be calculated at the village level, but there are fewer images at the village scale. This paper chooses to use townships as the unit for calculating wealth. Typically, a township contains an average of more than 20 villages. All wealth-related variables are aggregated at the township level.

In the dimension of averages, there are four indicators: floor height, base area, quality of agricultural housing, and brightness value of lights. In the proportion dimension, there are five indicators: car rate, motorcycle rate, air conditioning rate, rate of tiled exterior walls, and rate of exposed exterior walls. In the scale dimension, the total number of houses in each township is obtained using spatial statistics. The proportion indicators are calculated according to equation (2):

$$R_v = \frac{\sum_i^N D_i}{N} \qquad (2)$$

where $R_v$ represents the proportion for the corresponding variable. $D_i$ is a dummy variable that equals 1 when the related element is successfully detected and 0 when it is not detected. $N$ is the number of houses or images at the township level. Among them, the rates of tiled exterior walls, exposed exterior walls, and air conditioners are calculated based on the number of houses. Since vehicles cannot be identified as belonging to any particular agricultural house, the car rate and motorcycle rate are calculated based on the number of images.

### 2.2.3 Random Forest for Establishing the Regression Relationship Between Wealth Features and Labels

This paper uses random forest to train a regression model for inferring family wealth from image feature indicators. Random forest is a classifier that includes multiple decision trees, where the final prediction result depends on the mode of the outputs from each decision tree. Random forests have the following advantages: they do not require normalization of the data, can perform nonlinear prediction tasks, and can eliminate the interference of missing values, thereby ensuring a broader range of prediction. In addition, sub-indexes are often non-linearly distributed, such as floor height, car rate, etc., which are mainly long-tail distributions. Random forests can avoid estimation bias caused by non-normality of the data. Therefore, this paper selects random forest as the machine learning model.

This paper uses ten-fold cross-validation for model training. Specifically, the 1,678 townships are randomly divided into 10 groups, each group is used as the validation group, and the other 9 groups as the training group for training and validation. By averaging the evaluation of multiple models, a more robust model with generalizable predictive effects is obtained. During the model training process, some parameter indicators are set. The number of decision trees is set to 100 (ntree=100), which means the random forest model contains 100 decision trees. The number of features randomly sampled at each node is set to 1 (nfeature=1), indicating that each decision tree randomly selects 1 feature to evaluate when splitting a node.

## 3 Result

### 3.1 Comprehensive Evaluation of Wealth Index Prediction

Our study employed principal component analysis (PCA) on 13 wealth-related sub-indexes from 1,678 townships to construct a wealth index. The first principal component, accounting for 44% of the variance, was identified as the primary indicator of rural household wealth. Notably, key indicators such as flush toilet rate (0.86), bathroom rate (0.90), cooling facility rate (0.82), family income level (0.69), and summer monthly electricity bill (0.81) closely correlated with the wealth index, highlighting their potential to discern economic disparities (Table S6).

However, certain indicators like the building base area (0.33) and tap water supply rate (0.21) showed weaker correlations. This variation underscores the complex relationship between structural community features and individual wealth, as these measures may be influenced by local policy or widespread access, respectively.

Subsequently, we tested the accuracy of a random forest model in predicting these wealth indices. The model showed excellent predictive performance for the composite wealth index (correlation coefficient of 0.85, RMSE of 0.55), suggesting that the wealth index effectively encapsulates crucial wealth-differentiating information. The model's predictive power was particularly strong in the asset domain, with a high correlation for the prediction of floor height (0.89), reflecting the detailed image interpretation capabilities of our approach.

Family income predictions were less correlated (0.66), potentially due to the method of categorization based on income thresholds, which may not capture the full spectrum of income variability. Meanwhile, consumption indicators such as electricity bills, representing continuous numerical data, were predicted with high accuracy (0.81), implying that such consumption-based measures might offer a reliable representation of wealth levels.

The random forest models' overall capacity to predict both the composite wealth index and individual sub-indexes affirm the viability of employing machine learning techniques with geospatial data for detailed socioeconomic studies. Complete statistical analyses, including all model performance indicators, are provided in the S2 of SM.

## 3.2 Spatial Distribution and Characteristics of Rural Household Wealth

Analysis of the composite wealth index for rural households reveals a national average of -0.06, reflecting relative economic status rather than deficits, with pronounced wealth disparities across regions. Higher wealth concentrations appear along the Yangtze River and southeastern coastal regions, while a bimodal distribution pattern suggests two distinct clusters of wealth levels in the country (Figure 2).

As shown in Table 1, the average quality score of rural housing in national townships is 5.78. The average floor height is 1.59 floors, of which 85.6% of townships have floor heights of less than 2 floors, indicating that rural housing in China mainly consists of 1 or 2 floors. The difference in rural housing quality and floor height is more pronounced in the north-south direction, with larger building areas in the mid-latitude regions. The average building base area is 107 square meters, which is close to the 110 square meters reported in the questionnaire statistics. An average of 24% of rural houses have air conditioning facilities, with higher air conditioning rates mainly concentrated around the Beijing area and along the Yangtze River. 34% of rural houses have tiled exterior walls, while 11% have exposed exterior walls. Exposed exterior walls are more common in the western regions. The distribution of tiling rates is relatively less distinct. In terms of transportation tools, cars were captured in 14% of the images, while motorcycles accounted for only 11%. Due to differences in photography time and location, village street view images may not fully capture the number of cars or motorcycles, but they still reflect the close relationship between them and wealth. The proportion of cars and motorcycles is higher in the northern regions of Inner Mongolia.

*Table 1 Descriptive Statistical Features of Wealth Elements.*

| Dimension | Feature Indicator | Average | Maximum | Minimum | Variance |
|---|---|---|---|---|---|
| Wealth | Composite Wealth Index | -0.06 | 2.27 | -2.52 | 0.82 |
| Housing | Number of Houses | 13,072 | 161,718 | 1,524 | 10,828.63 |
|  | Building Base Area | 107 | 181 | 0 | 18.90 |
|  | Rate of Old-Style Agricultural Houses | 0.26 | 1.00 | 0 | 0.18 |
|  | Floor Height | 1.59 | 6.93 | 1 | 0.42 |
|  | Quality of Agricultural Housing | 5.78 | 7.10 | 2.06 | 0.33 |
|  | Rate of Tiled Exterior Walls | 0.34 | 1.00 | 0 | 0.15 |
|  | Rate of Exposed Exterior Walls | 0.11 | 0.86 | 0 | 0.10 |
| Facilities | Air Conditioning Rate | 0.24 | 8.48 | 0 | 0.30 |
|  | Car Rate | 0.14 | 1.00 | 0 | 0.10 |
|  | Motorcycle Rate | 0.11 | 1.00 | 0 | 0.08 |

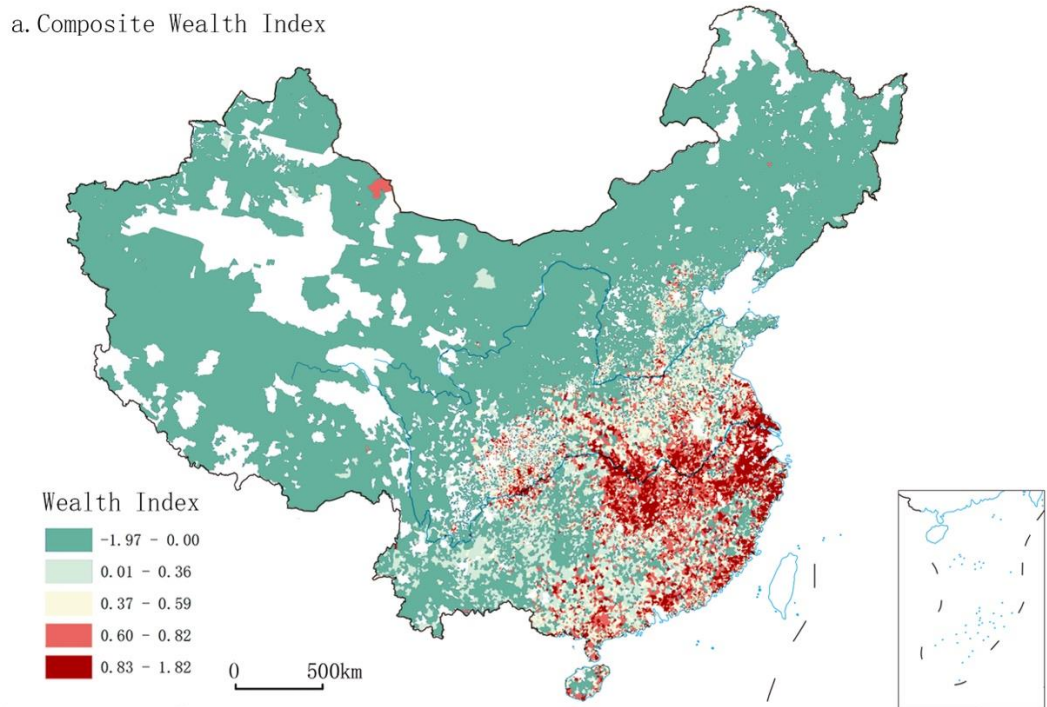
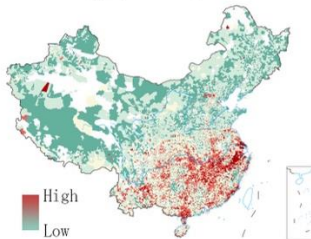
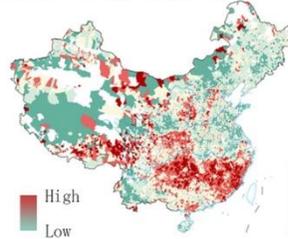
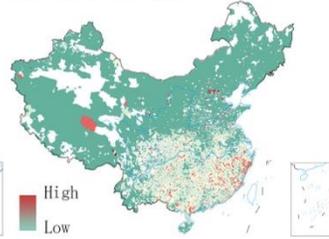
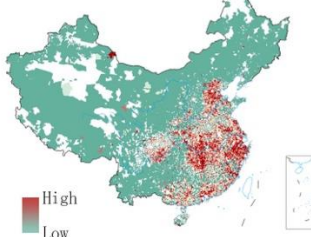
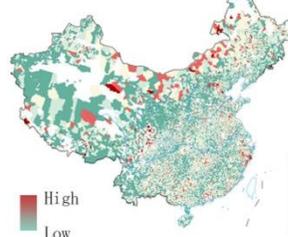
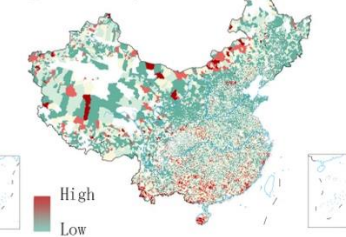
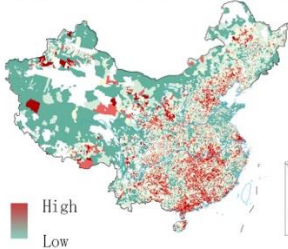
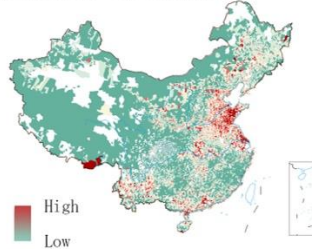
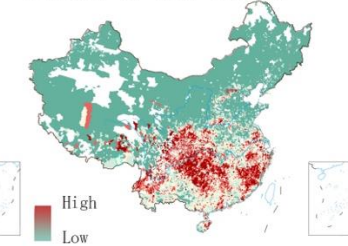

*Figure 2 Composite Wealth Index. Note: Based on the standard map produced by the Ministry of Natural Resources standard map service website, GS (2016) No. 1569, the map boundaries have not been modified.*

China's rural household wealth is markedly heterogeneous. County-level analyses show a bimodal wealth distribution, suggesting polarization into higher and lower wealth zones. Wealthier areas, particularly in the

Yangtze River and southeastern coastal regions, contrast starkly with poorer counterparts in the northwest, northeast, and Tibetan Plateau.

Geographic disparities are pronounced, with wealth typically greater in the south than in the north and in the east more than in the west. The demarcating Hu Line and Qinling Huaihe Line reflect these disparities, with wealth indices on their east and south sides considerably higher than on their west and north. Within China's nine major agricultural zones, wealth peaks in the Yangtze River's fertile basin and subsides in the remote Qinghai-Tibet, Northeast, and Inner Mongolia regions.

A granular view, informed by intelligently interpreted features, confirms southern regions surpass the north in wealth indices, with homes here averaging higher floor counts and larger base areas Table 2. Modern construction materials, air conditioning, and vehicle ownership rates further delineate the divide, showcasing the south's deeper investment in housing and development.

China's wealthiest rural counties, representing the top 10% in household wealth index, cluster along the Yangtze River and southeast coast, forming economic belts enriched by decades of progressive reform (Figure 3). Notably, these regions harness the advantages of special economic zones like Xiamen and Shenzhen, blending traditional agriculture with modern enterprises. Models of urbanization, such as the Wenzhou and Putian models, exemplify the seamless integration of rural spaces into the economic fabric without uprooting communities.

The fecund basins around Dongting, Tai, and Poyang Lakes bolster this wealth, supporting rice agriculture that thrives on surplus for both local sustenance and nationwide supply. This prosperity has allowed for significant reinvestment in rural infrastructure and housing, amplifying the wealth of these areas.

Conversely, the bottom 10% of counties, predominantly west of the Hu Line and in the northeast, struggle with geographic and infrastructural barriers. Over half of these counties lie in border areas with limited access to national economic streams and infrastructural support, hampered by rugged plateau terrains and underdeveloped agriculture. The disparity underscores China's diverse economic landscape and highlights the need for targeted development strategies in its remote rural regions.

*Table 2 Differences in Wealth Variables Between Rural Areas in the North and South*

| Dimension | Feature Indicator | Hu Line | Qinling-Huaihe Line |
|---|---|---|---|
| | | East Side | West Side |
| **Wealth** | Composite Wealth Index | 0.07 | -0.99 |
| **Housing** | Number of Houses | 14,475 | 7,740 |
| | Building Base Area | 107.56 | 95.96 |
| | Rate of Old-Style Agricultural Houses | 0.25 | 0.15 |
| | Floor Height | 1.63 | 1.29 |
| | Quality of Agricultural Housing | 5.80 | 5.56 |
| | Rate of Tiled Exterior Walls | 0.34 | 0.26 |
| | Rate of Exposed Exterior Walls | 0.11 | 0.13 |
| **Facilities & Equipment** | Air Conditioning Rate | 0.28 | 0.06 |
| | Car Rate | 0.16 | 0.12 |
| | Motorcycle Rate | 0.12 | 0.09 |

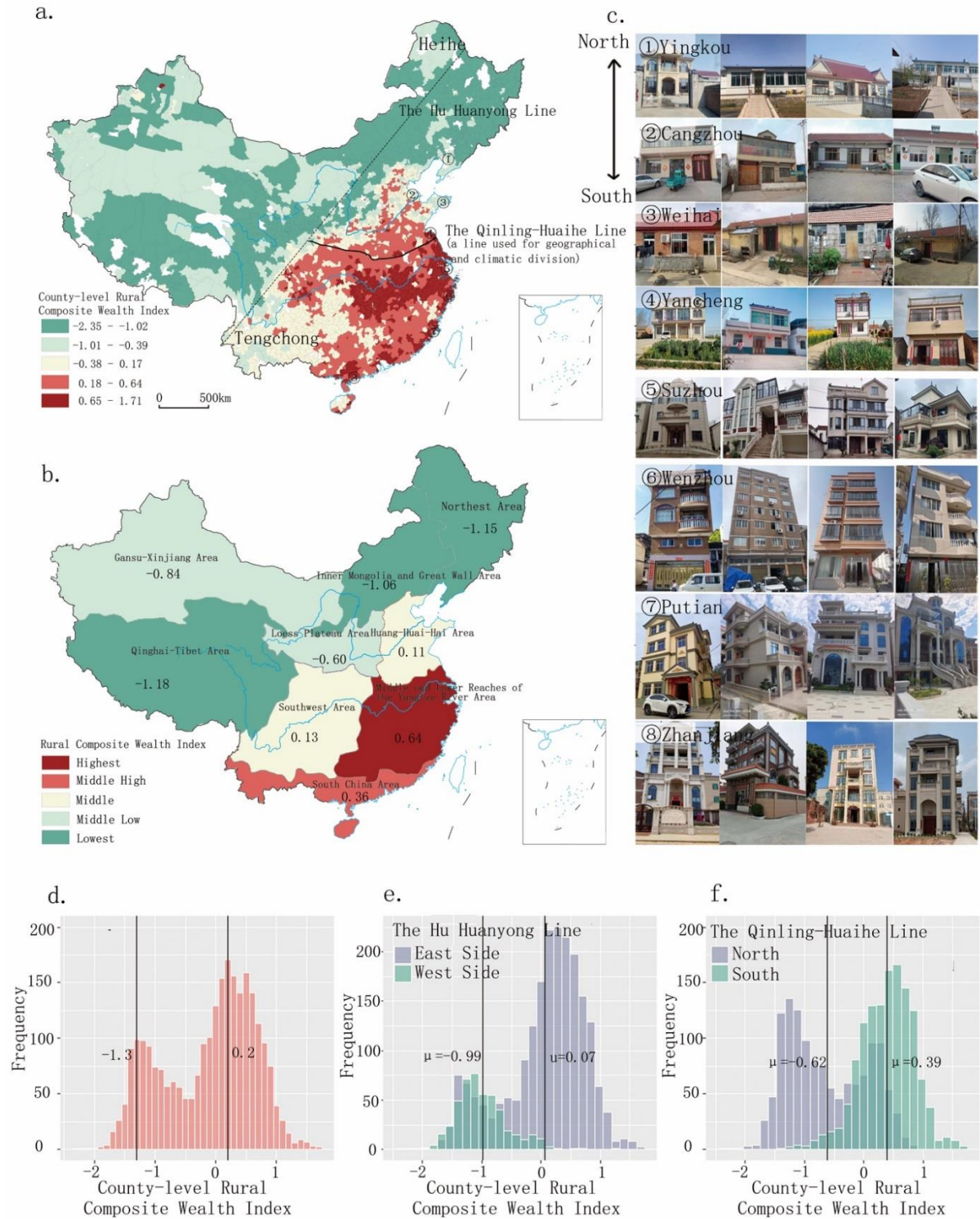

*Figure 3 Differentiation map of the Composite Wealth Index. a) is the county-level composite wealth index; b) is the composite wealth index of the nine major agricultural zoning areas; c) is the village street view of different wealth levels from north to south; d) is the frequency histogram of the composite wealth index at the county level; e) is the frequency histogram grouped by the east and west sides of the Hu Line; f) is the histogram grouped by the sides of the Qinling Huaihe Line. Based on the standard map from the Natural Resources Department's standard map service website, GS (2016) No. 1569, the base map boundaries have not been modified.*

# 4 Discussion

Rural family wealth is crucial to sustainable development goals and is closely related to the realization of a Chinese-style modernization of shared prosperity. Among many assets of wealth, housing is the most important part. Rural housing is the most important asset in the life cycle of a farmer and is also a major representation of income, with supplementary facilities such as air conditioning and cars closely related to rural housing. Understanding the current level of rural family wealth centered around rural housing, analyzing the spatial distribution patterns and main characteristics of rural housing, and formulating policies to promote rural development are of great significance. Rural family wealth data can not only measure the level of rural development but also provide an important basis for investment construction and transfer payments in rural areas, thereby offering key support for post-policy implementation assessment and forming an important foundation for rural research, policy formulation, and implementation. Therefore, it is urgent to draw up a comprehensive, large-scale, and finely-grained map of Chinese rural family wealth to explore the key drivers of wealth spatial differentiation and thereby promote the overall growth of rural family wealth.

This paper proposes a technical framework for identifying and evaluating the wealth elements of Chinese rural families, centered on the intelligent interpretation of rural housing, integrating "sky" remote sensing imagery with "ground" village street view imagery, through a series of deep learning technologies. Under this overall technical framework, for the first time, this paper has drawn a comprehensive index map of rural family wealth at the national township scale that encompasses wealth-related elements and analyzed the regional differences in rural family wealth.

The results show that the correlation coefficient between the predicted and actual values of the composite index of rural family wealth is as high as 0.85. In terms of rural housing assets, the correlation coefficients of wealth indicators such as rural housing height, independent bathroom rate, flush toilet rate, and cooling facility rate also exceeded 0.8; in terms of consumption, the correlation coefficient for summer monthly electricity fees was 0.81. In terms of income, the prediction effect for family income was slightly insufficient, with a correlation coefficient of 0.66. The paper then predicted images from 185,000 villages across 30,667 townships nationwide, combined with the rural housing database to form 10 indicators for model feature input, and successfully drew a national-scale, township-level map of the comprehensive index of Chinese rural family wealth, thus precisely revealing the regional differences in rural family wealth across a wide range. Overall, the distribution of rural family wealth in China exhibits significant polarization, with clear wealth differentiation features of "high in the east, low in the west, high in the south, low in the north." The counties ranking in the top 10% of the composite index of rural family wealth are mainly distributed in the middle and lower reaches of the Yangtze River and the southeastern coastal regions, while those ranking in the last 10% are mainly located west of the Hu Line.

This paper still has shortcomings in terms of spatial granularity. Villages are the basic units of national governance and the micro spatial entities of interaction between the state and the countryside. Rural revitalization takes the village as the main unit. The current map of the family wealth index has not reached the village level, which inevitably has its shortcomings. In terms of remote sensing, remote sensing interpretation can directly reach every rural house, but street view images are rarer. Currently, there are only about five images of rural houses per village in this paper, which may lead to a problem of insufficient sample size, hence the choice to use townships as the spatial statistical unit. In the future, as village street view crowdsourcing platforms are promoted and popularized, villages may collect more images of rural houses, and at that time, it will be possible to explore the evolution of rural family wealth at the village scale.